# FABRICATION, STUDY OF OPTICAL PROPERTIES AND STRUCTURE OF MOST STABLE $(CdP_2)_n$ NANOCLUSTERS

O.A. Yeshchenko*, I.M. Dmitruk, S.V. Koryakov, M.P. Galak

*Physics Department, National Taras Shevchenko Kyiv University,*
*2, build #2, Akademik Glushkov prosp., 03022 Kyiv, Ukraine*

$CdP_2$ nanoclusters were fabricated by incorporation into pores of zeolite Na-X and by laser ablation. Absorption and photoluminescence (PL) spectra of $CdP_2$ nanoclusters in zeolite were measured at the temperatures of 4.2, 77 and 293 K. Both absorption and PL spectra consist of two bands blue shifted with respect to bulk crystal. We performed the calculations aimed to find the most stable clusters in the size region up to size of the zeolite Na-X supercage. The most stable clusters are $(CdP_2)_6$ and $(CdP_2)_8$ with binding energies of 9.30 eV and 10.10 eV per $(CdP_2)_1$ formula unit respectively. Therefore, we attributed two bands observed in absorption and PL spectra to these stable clusters. The Raman spectrum of $CdP_2$ clusters in zeolite was explained to be originated from $(CdP_2)_6$ and $(CdP_2)_8$ clusters as well. The PL spectrum of $CdP_2$ clusters produced by laser ablation consists of the asymmetric band with low-energy tail that has been attributed to emission of both $(CdP_2)_8$ cluster and $CdP_2$ microcrystals.

## 1. Introduction

There exist many different methods for fabrication of the semiconductor nanoparticles, e.g. fabrication of nanoparticles in solutions [1], glasses [2] or polymers [3]. However, it is not easy to control the size distribution of small nanoparticles with countable number of atoms (so called clusters) using these methods. Matrix method based on the incorporation of materials into the 3D regular system of voids and channels of zeolites crystals could be one of the possible solutions [4,5]. Moreover, the subnanometer and nanometer clusters are very interesting as they are intermediate between the molecules and the typical nanocrystals. Usually, the structure of nanoclusters is different from the structure of nanocrystals, which resembles the structure of bulk crystals. As a rule, the methods of calculation of the structure of electronic states of nanocrystals that are based on the effective mass approximation are not applicable for clusters. Thus, nanoclusters are very interesting objects as their structure, electronic and vibrational properties are quite different from the crystalline nanoparticles. Zeolites provide the opportunity to obtain extremely small clusters in the pores with diameter up to 15 Å. Zeolites are crystalline alumosilicates with cavities which diameter can vary in the range from 7 to15 Å. It depends on the type of alumosilicate framework, ratio Si/Al, origin of ion-exchanged cations, which stabilise negative charge of framework, etc. Zeolite Na-X, which has been used in the present work has Si/Al ratio equal 1, Fd3m symmetry and two types of cages: one is sodalite cage – truncated octahedron with diameter 8 Å and supercage, which is formed by the connection of sodalites in diamond-like structure with the diameter of about 13 Å [6]. All cages are interconnected by shared small windows and arranged regularly. Thus, the cages can be used for fabrication of small semiconductor nanoclusters.

---

* Corresponding author: O.A.Yeshchenko.
Tel.: +380-44-2664587;   Fax: +380-44-2664036;   E-mail: yes@univ.kiev.ua

Laser ablation (LA) is a well-known method to produce nanoclusters by ablating material from a solid target [7]. LA usually is performed in vacuum, or sometimes in inert gas such as Ar or more reactive gases such as ammonia or nitrogen. Recently a new variation of LA has been reported whereby the target is immersed in a liquid medium, and the laser beam is focused through the liquid onto the target surface [8]. LA technique has been used to produce nanoclusters of semiconductors (see e.g. Refs. [9,10]) and metals (see e.g. Ref.. [11]).

The nanoclusters of II-V semiconductors are studied rather poorly. As we know there are several works on $Cd_3P_2$ nanoclusters fabricated by wet chemistry methods [12,13] and by thermolysis [14] and alcoholysis of organometallic species [15]. As well, in our recent work [16] we have reported the fabrication and study of the optical properties of the nanoclusters of another II-V semiconductor ($ZnP_2$) incorporated into zeolite Na-X matrix. The present paper is the first study of the nanoclusters of another II-V semiconductor: cadmium diphosphide ($CdP_2$). Wet chemistry methods seem to us not to be suitable for production of ultrasmall II-V nanoclusters due to their high reactivity in water. It is hard to expect their high stability in glass melt as well. Thus, incorporation into zeolite cages and production by laser ablation seem to us to be ones of the most suitable methods for II-V semiconductor nanoclusters fabrication.

Quantum confinement of charge carriers in nanoclusters leads to new effects in their optical properties. Those are the blue shift of exciton spectral lines originating from the increase of the kinetic energy of charge carriers and the increase of the oscillator strength per unit volume [17,18]. These effects are quite remarkable when the radius of the nanoparticle is comparable with Bohr radius of exciton in bulk crystal. Incorporation into zeolite pores and laser ablation are quite promising methods for fabrication of small nanoclusters in which these effects are considerable to be studied.

Bulk $CdP_2$ crystal is the indirect-gap semiconductor. The symmetry of its lattice is characterised by the space symmetry group $D_4^4$ for right-rotating and $D_4^8$ for left-rotating modification (tetragonal syngony). The $CdP_2$ energy gap is 2.155 eV [19]. Since the bulk crystal is red at the temperature of 77 K, the blue shifted exciton lines of $CdP_2$ nanoclusters are expected to be in visible or near UV spectral region.

## 2. Technology of fabrication of $CdP_2$ nanoclusters. Experimental procedures

For the fabrication of $CdP_2$ nanoclusters we used high purity $CdP_2$ bulk crystals and synthetic zeolite of Na-X type. The framework of zeolite Na-X consists of sodalite cages and supercages with the inner diameters of 8 and 13 Å, respectively. $CdP_2$ nanoclusters are too large to be incorporated into small sodalite cage due to existence of many Na cations. Therefore, it is naturally to assume that only the supercages can be the hosts for the nanoclusters. Zeolite and $CdP_2$ crystals were dehydrated in quartz ampoule in vacuum about $2\times10^{-5}$ mm Hg for 1 h at 400°C. Then ampoule was sealed. We used 100 mm length ampoule for space separation of semiconductor source and zeolite. The fabrication of samples was carried out in two stages. At the first stage (see Fig.1) $CdP_2$ was incorporated into the zeolite matrix through the vapour phase at 769°C in source region and 760°C in zeolite region for 100 h. At the second stage, the inverted temperature gradient was applied: 759°C in source region and 763°C in zeolite region. The duration of the second stage was 40 h. The cooling of ampoule we carried out gradually with above mentioned inverted temperature gradient. The stability of structure of lattice of zeolite single crystals was controlled by XRD method. The control showed that at above mentioned temperatures the zeolite lattice structure was stable, i.e. semiconductor nanoclusters were incorporated into the single crystal zeolite matrix. The loading level of $CdP_2$ into zeolite was 11.7 wt% that corresponds to average value of 1.2 formula units per zeolite supercage.

During the optical measurements the samples were as in vacuum in quartz ampoule and in air. A tungsten-halogen incandescent lamp was used as a light source for the diffuse reflection measurements. An $Ar^{++}$ laser with wavelength 351.1 nm was used for the excitation of the luminescence, and $Ar^+$ laser with wavelength 514.5 nm – in Raman experiments. The absorption spectra of the nanoclusters were obtained from the diffuse reflection spectra by conversion with Kubelka-Munk function $K(\hbar\omega)=[1-R(\hbar\omega)]^2/2R(\hbar\omega)$, where $R(\hbar\omega)$ is the diffuse reflectance normalised by unity at the region of no absorption.

For ablation the pulsed Cu laser ($\lambda$=5782 Å) was used. The pulse intensity of the focused laser beam was about 1.5 MW/cm$^2$, pulse duration was of 20 ns at a repetition rate of 10 kHz. The beam was focused on the surface of target to a spot-size diameter of approximately 0.5 mm. During the ablation the target (bulk $CdP_2$ crystal of area 12 mm$^2$) was dipped into the liquid nitrogen. The produced by ablation nanoclusters of $CdP_2$ were deposited on quartz plate that was positioned on the distance of about 0.2 mm from the irradiated surface of the target crystal. The run time of ablation was 30 min. The photoluminescence spectrum was measured from the deposited film by cw Ar$^{++}$ laser ($\lambda$ =351.1 nm).

### 3. Structure and optical properties of CdP$_2$ nanoclusters

Diffuse reflection (DR) and photoluminescence (PL) spectra of the CdP$_2$ nanoclusters incorporated into the 13 Å supercages of zeolite Na-X were measured at room (290 K), liquid nitrogen (77 K) and liquid helium (4.2 K) temperatures. Then, the DR spectrum was converted to absorption one by the Kubelka-Munk method described above. Within the accuracy of determination of bands spectral positions we did not observe noticeable change of both absorption and PL spectra with temperature. Optical spectra of the clusters are the same both for the samples placed in vacuum in quartz ampoule and for ones placed in air. This is an evidence of the stability of CdP$_2$ clusters in pores of zeolite placed in air. The absorption spectrum obtained by Kubelka-Munk method is presented in Fig. 2. The spectrum demonstrates two-band structure. The spectral positions of the respective bands signed as B$_1$ and B$_2$ are presented in Table 1. Both the bands are blue shifted (Table 1) with respect to spectrum of bulk crystal. Let us note that the blue shift of the high-energy absorption band for CdP$_2$ clusters in Na-X zeolite is larger than respective one for ZnP$_2$ clusters in the same zeolite [16]: 1.151 eV for CdP$_2$ and 0.808 eV for ZnP$_2$. The observed blue shift allows us to attribute these bands to the absorption into the first electronic excited state of CdP$_2$ nanoclusters incorporated into supercages of the zeolite. The photoluminescence spectrum of CdP$_2$ clusters in zeolite (Fig. 3(*a*)) shows the same structure as the absorption one, i.e. PL spectrum consists of the corresponding two $B'_1$ and $B'_2$ bands. Their spectral positions are presented in Table 1. PL bands of nanoclusters are blue shifted from the spectrum of bulk crystal as well. The observed blue shift of the absorption and luminescence bands is the result of the quantum confinement of electrons and holes in CdP$_2$ nanoclusters.

It is often observed that nanoclusters with certain number of atoms are characterised by elevated stability (ultrastable nanoclusters) and are more abundant in the sample. This effect is well known for the nanoclusters of different types, e.g. for C [20], Ar [21], Na [21], and for nanoclusters of II-VI semiconductors [22,23]. Our first-principles calculations [16] have shown that such stable nanoclusters exist for ZnP$_2$. Those are (ZnP$_2$)$_6$ and (ZnP$_2$)$_8$ with binding energies 5.16 eV and 6.42 eV per formula unit respectively. Thus, it would be naturally to assume that similarly to ZnP$_2$ the respective stable (CdP$_2$)$_n$ nanoclusters exists for CdP$_2$ as well. We performed the calculations aimed to find such stable (CdP$_2$)$_n$ clusters. Initially, we performed the geometry optimization of the structure of clusters by molecular mechanics method. Then, we performed the *ab initio* calculation of the ground state energy of the clusters with optimized structure. The results are presented in Fig.4. It is seen from the figure that CdP$_2$ molecule does not exist as it is unbound. The clusters with even *n* have higher binding energy than clusters with odd n. Likely to ZnP$_2$ the (CdP$_2$)$_n$ clusters with $n = 6$ and 8 are the most stable. The (CdP$_2$)$_6$ cluster is characterised by the binding energy 9.30 eV per formula unit, and the (CdP$_2$)$_8$ one – 10.10 eV per formula unit, i.e. the (CdP$_2$)$_n$ clusters are bound stronger than the respective (ZnP$_2$)$_n$ ones. The maximum diameter of (CdP$_2$)$_6$ cluster is 9.38 Å, and the maximum diameter of (CdP$_2$)$_8$ is 9.48 Å. Here and everywhere in the article, maximum diameter means the distance between the centers of outermost atoms of cluster. The structure of these ultrastable clusters is presented in Fig.5. It is seen that as it can be expected the structures of (CdP$_2$)$_6$ and (CdP$_2$)$_8$ clusters are the same as the structures of the respective (ZnP$_2$)$_n$ clusters. (CdP$_2$)$_6$ looks like some nanotube, and (CdP$_2$)$_8$ looks like some nanodisk. Similar to the structure of bulk CdP$_2$ crystal both these clusters have six-membered rings of atoms. It is naturally to assume that some other stable (CdP$_2$)$_n$ clusters exist as well. One of those is (CdP$_2$)$_{10}$ cluster with maximum diameter of 11.57 Å and binding energy of 9.80 eV per formula unit. However, our estimation of the diameter of largest (CdP$_2$)$_n$ cluster that might be placed in zeolite Na-X supercage is about 9.5 Å. Our estimations consider the covalent radii of atoms of cluster and of inner atoms of zeolite supercage. Therefore, the clusters with $n > 8$ can not be incorporated into supercages of Na-X zeolite. Since the (CdP$_2$)$_6$ and (CdP$_2$)$_8$ clusters are the most stable, it is quite reasonable to assume that these clusters are the most abundant. Thus, B$_1$ and $B'_1$, B$_2$ and $B'_2$ bands can be attributed to originate from above mentioned stable (CdP$_2$)$_6$ and (CdP$_2$)$_8$ nanoclusters incorporated into the supercages of zeolite matrix.

One can see from the Table 1 that the luminescence bands have the Stokes shift from the absorption ones. The values of this shift are large enough: 0.774 eV for $B'_1$ band and 0.484eV eV for $B'_2$ one. These values are considerably larger than ones for ZnP$_2$ nanoclusters in the same Na-X zeolite (0.078-0.135 eV: see Ref. [16]). Stokes shift is well known both in the molecular spectroscopy and in the spectroscopy of nanoclusters. It is known that this kind of Stokes shift (so-called Frank-Condon shift) is due to vibrational relaxation of the excited molecule or nanoparticle to the ground state. The theory of Frank-Condon shift in nanoclusters was developed in Ref. [24] where the first-principle calculations of excited-state relaxations in nanoclusters were performed. As it is shown in Ref. [24], for small nanoclusters the Stokes shift is the Frank-Condon one, which is the result of the vibrational relaxation of the nanoparticle in the excited electronic state. The considerable values of Stokes shift in (CdP$_2$)$_n$ clusters mean the substantial role of vibrational relaxation in excited nanoparticles.

To check our above assumption of the origin of the absorption and PL bands of the CdP$_2$ incorporated into zeolite we performed Raman study of the samples. The obtained Raman spectrum is shown in Fig.6. The Raman spectrum consists of a single asymmetric rather wide band centered at 266 cm$^{-1}$ with half-width of 15 cm$^{-1}$. To check our above assumption of the formation of (CdP$_2$)$_6$ and

$(CdP_2)_8$ clusters in zeolite supercages we performed semi-empirical calculation of the frequencies of vibrations of these clusters. Calculations showed that $(CdP_2)_6$ and $(CdP_2)_8$ clusters would have the Raman active group of vibrational normal modes with frequencies in the range 257.5 – 279.5 cm$^{-1}$. Therefore, the Raman spectrum proves our assumption of the formation of $(CdP_2)_6$ and $(CdP_2)_8$ clusters in zeolite pores.

Additionally to clusters in zeolite pores, we fabricated the $CdP_2$ nanoclusters by laser ablation technique described above. The obtained PL spectrum of $CdP_2$ clusters is shown in Fig.3(b). The spectrum consists of an asymmetric band with clear low-energy tail. One can see from the figure that two bands contribute to this asymmetric band. The spectral position of the maximum of first band, signed as $B_2''$, coincides with the position of $B_2'$ band of luminescence spectrum of $CdP_2$ clusters in zeolite. This fact and the proximity of values of the half-widths of $B_2''$ and $B_2'$ bands (0.24 eV and 0.17 eV correspondingly) allows us to assume the same origin of these PL bands, i.e. that both the bands originate from the emission from excited to ground electronic state of $(CdP_2)_8$ cluster. This our assumption seems to be quite reasonable as such cluster is the most stable. Correspondingly, $(CdP_2)_8$ clusters would be formed in prevalent quantities at the ablation. An effect of the prevalence of the most stable nanoclusters in mass spectra is well known for II-VI semiconductors (see e.g. Refs. [22,23]). Meanwhile, laser ablation is the method for production of analysed particles in mass spectrometry. It is naturally that other $CdP_2$ nanoclusters, besides $(CdP_2)_8$, would be formed at the ablation as well, but, probably, their quantity is quite small compared to quantity of $(CdP_2)_8$ ones. Therefore, these less abundant clusters would not give the considerable contribution to PL spectrum. The maximum of the second band, marked as $B_b$, is characterized by lower energy than the energy gap of the bulk crystal. Proceeding from this, we can assume that $B_b$ band is the result of the emission from the $CdP_2$ microcrystals that appear at the laser ablation of macrocrystal target. Since the size of the microcrystals is quite large with respect to the exciton Bohr radius of bulk crystal, microcrystals are the bulk ones. Accordingly, it would not exist any blue shift of the spectrum of these crystals. Therefore, the observed PL spectrum of the microcrystals is characterized by lower energies than energy gap of the bulk crystals.

**References**

[1] N. Kumbhojkar, V.V. Nikesh, A. Kshirsagar and S. Nahamuni, J. Appl. Phys. **88** (2000) 6260.
[2] L. Armelao, R. Bertoncello, E. Cattaruzza, S. Gialanella, S. Gross, G. Mattei, P. Mazzoldi and E. Tondello, J. Appl. Chem. **12**, (2002) 2401.
[3] L. Motte and M.P. Pileni, Appl. Surf. Sci., **164** (2000) 60.
[4] V.N. Bogomolov, Usp. Fiz. Nauk **124** (1978) 171 [Sov. Phys. Usp. **21** (1978) 77].

[5] G.D. Stucky and J.E. Mac Dougall, Science **247**, (1990) 669.

[6] A. Corma, Chem. Rev. **95** (1995) 559.

[7] M. D. Shirk and P. A. Molian, J. Laser Appl. **10** (1998) 18.

[8] A. V. Simakin, G. A. Shafeev, and E. N. Loubnin, Appl. Surf. Sci. **154** (2000) 405.

[9] A. Burnin and J.J. BelBruno, Chem. Phys. Lett. **362** (2002) 341.

[10] A.B. Hartanto, X. Ning, Y. Nakata, T. Okada, Appl. Phys. A **78** (2004) 299.

[11] S.S.I. Dolgaev, A.V. Simakin, V.V. Voronov, G.A. Shafeev, Appl. Surf. Sci. **186** (2002) 546.

[12] H. Weller, A. Fojtik and A. Henglein, Chem. Phys. Lett., **117** (1985) 485.

[13] X.-G. Zhao, J.-L. Shi, B. Hu, L.-X. Zhang and Z.-L. Hua, J. Mater. Chem., **13** (2003) 399.

[14] M. Green and P. O'Brien, Adv. Mater., **10** (1998) 527.

[15] M. Green and P. O'Brien, J. Mater. Chem., **9** (1999) 243.

[16] O.A.Yeshchenko, I.M.Dmitruk, S.V.Koryakov, I.P. Pundyk, and Yu.A.Barnakov, Solid State Commun. **133** (2005) 109.

[17] L. E. Brus, J. Phys. Chem. **90** (1986) 2555.

[18] B.O. Dabbousi, J. Rodriguez-Viejo, F.V. Mikulec, J.R. Heine, H. Mattoussi, R. Ober, K.F. Jensen and M.G. Bawendi, J. Phys. Chem. B **101** (1997) 9463.

[19] I.S. Gorban, V.A. Gubanov, M.V. Chukichev and Z.Z. Yanchuk, Fiz. Tekhn. Poluprovodn. **19** (1985) 1312.

[20] H.W. Kroto, J.R. Heath, S.C. O'Brien, R.F. Curl and R.E. Smalley, Nature **318** (1985) 162.

[21] J.-P. Connerade, A.V. Solovýov and W. Greiner, Europhys. News **33 (6)** (2002) 200.

[22]. A. Kasuya, R. Sivamohan, Yu.A. Barnakov, I.M. Dmitruk, T. Nirasawa, V.R. Romanyuk, V. Kumar, S.V. Mamykin, K. Tohji, B. Jeyadevan, K. Shinoda, T. Kudo, O. Terasaki, Z. Liu, R.V. Belosludov, V. Sundararajan and Y. Kawazoe, Nature Mater. 3 (2004) 99.

[23] A. Burnin and J.J. BelBruno, Chem. Phys. Lett. **362** (2002) 341.

[24] A. Franceschetti and S. T. Pantelides, Phys. Rev. B **68** (2003) 033313.

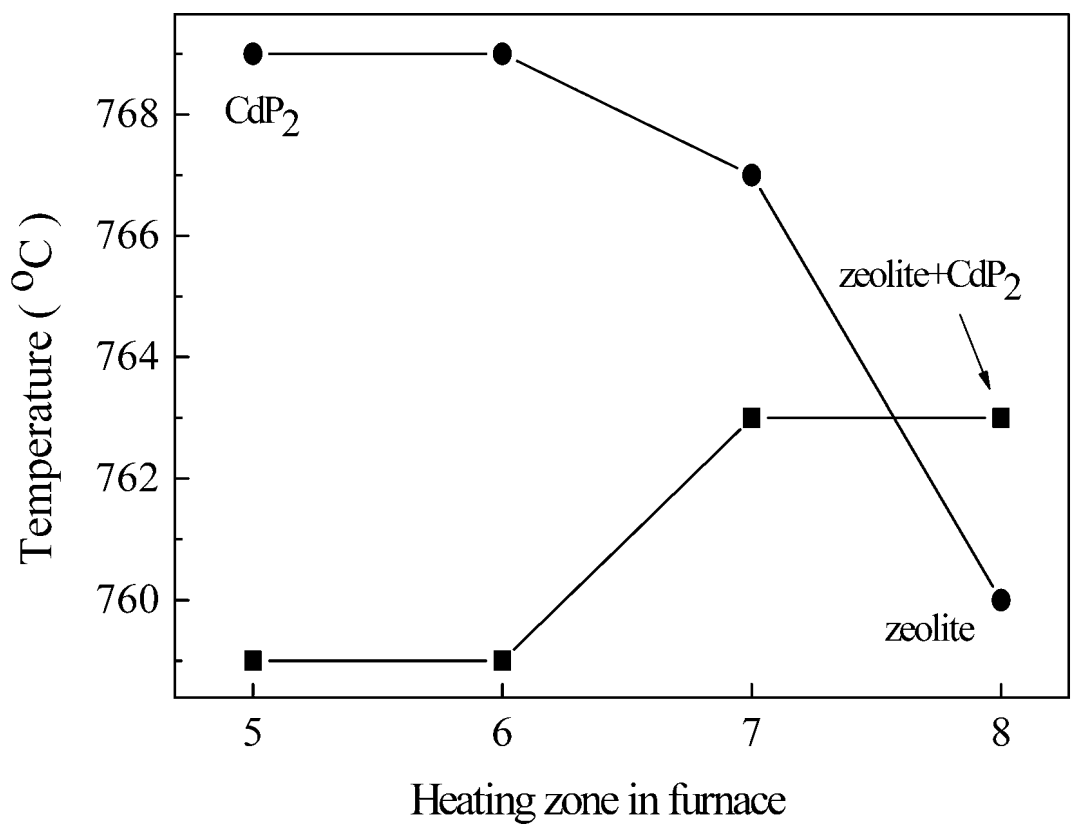

Fig.1.The temperature gradients used at first stage (connected filled circles) and at second stage (connected filled squares) of the fabrication of CdP$_2$ clusters in zeolite Na-X.

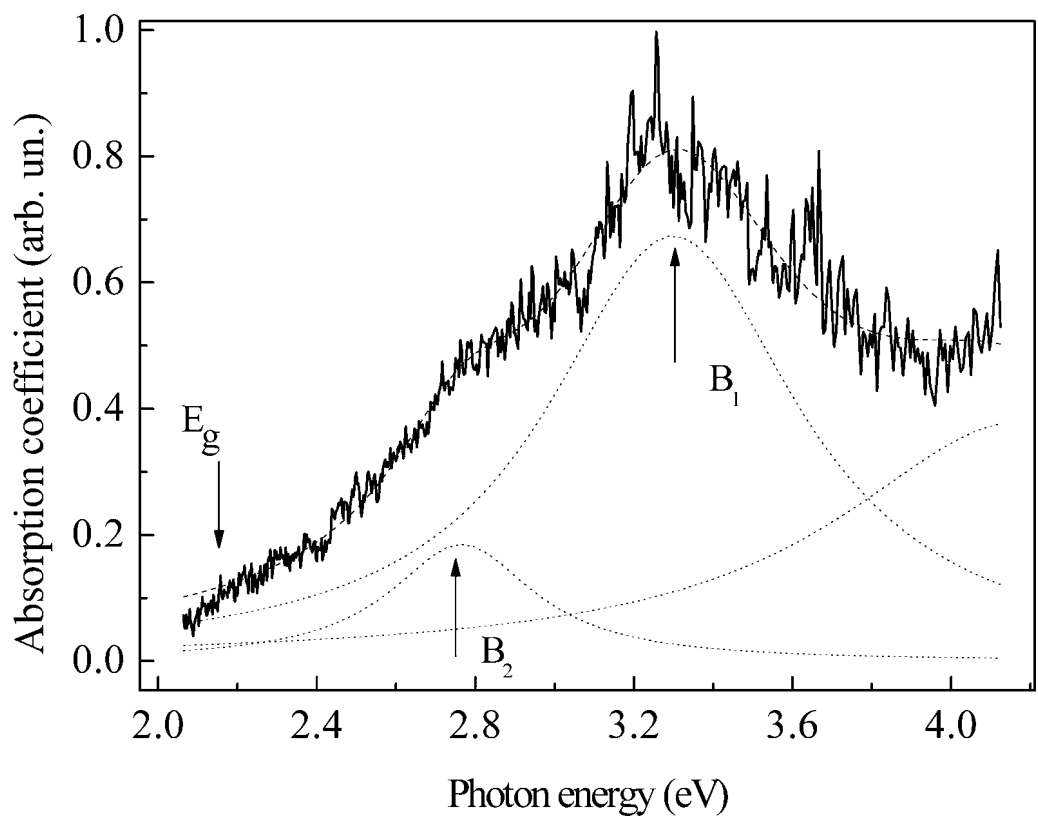

Fig. 2. The obtained by Kubelka-Munk method absorption spectrum of $CdP_2$ clusters in zeolite Na-X at the temperature of 77 K.

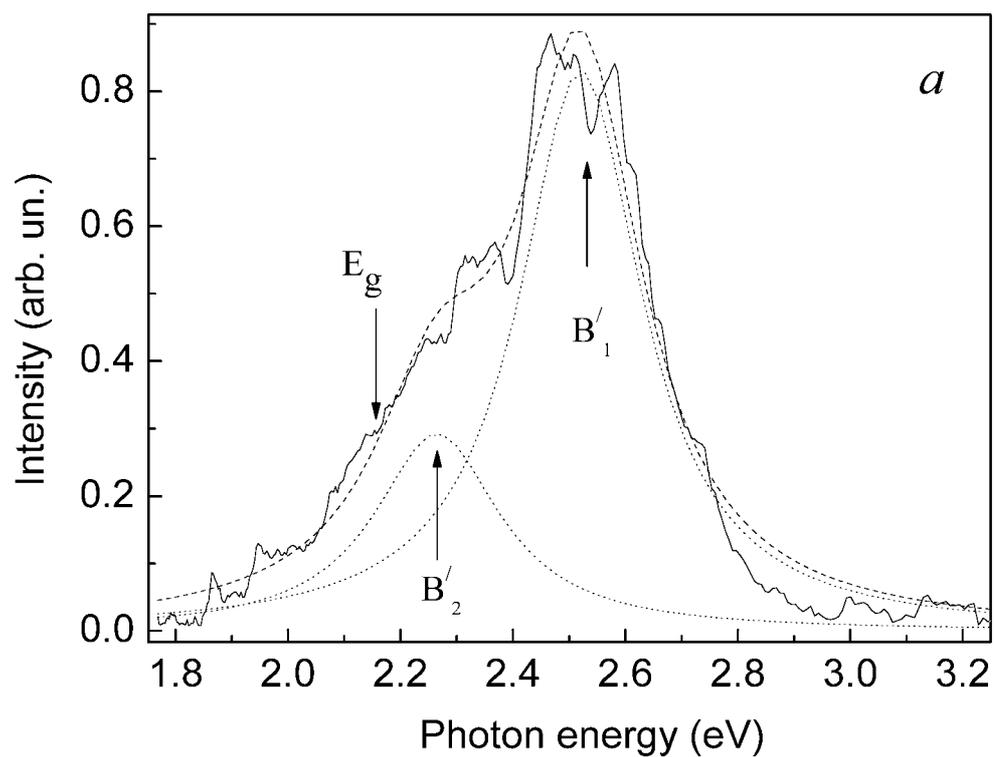

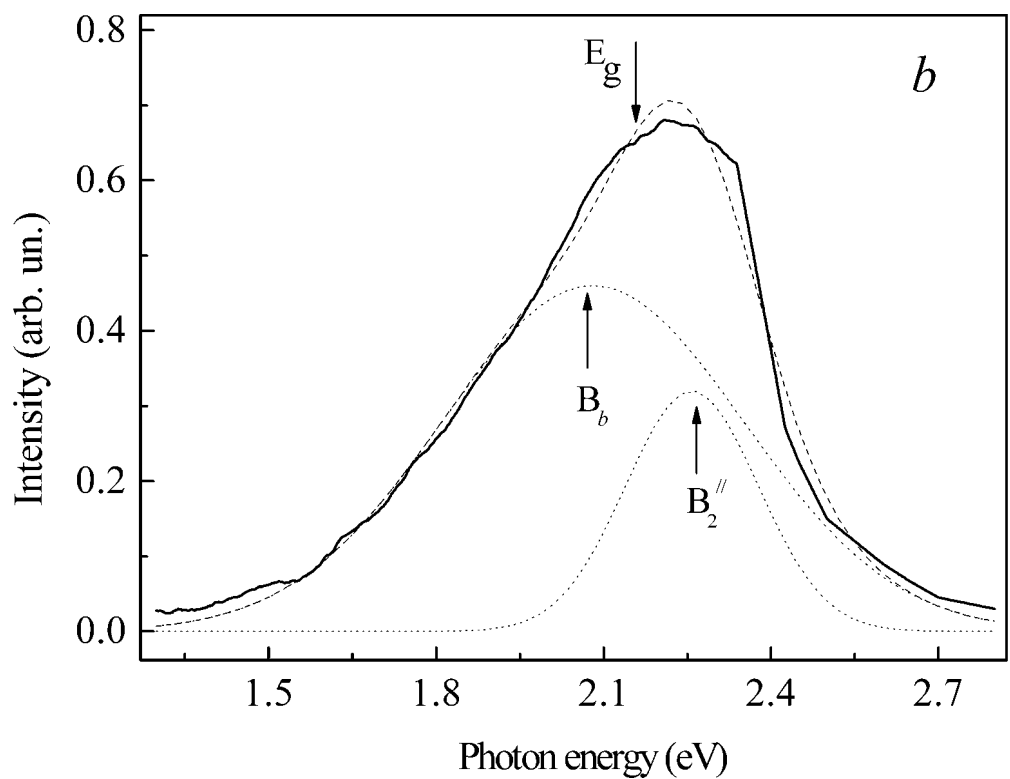

Fig.3. (a) – The photoluminescence spectrum of $CdP_2$ clusters in zeolite Na-X. (b) – The PL spectrum of $CdP_2$ particles produced by laser ablation at the temperature of 77 K.

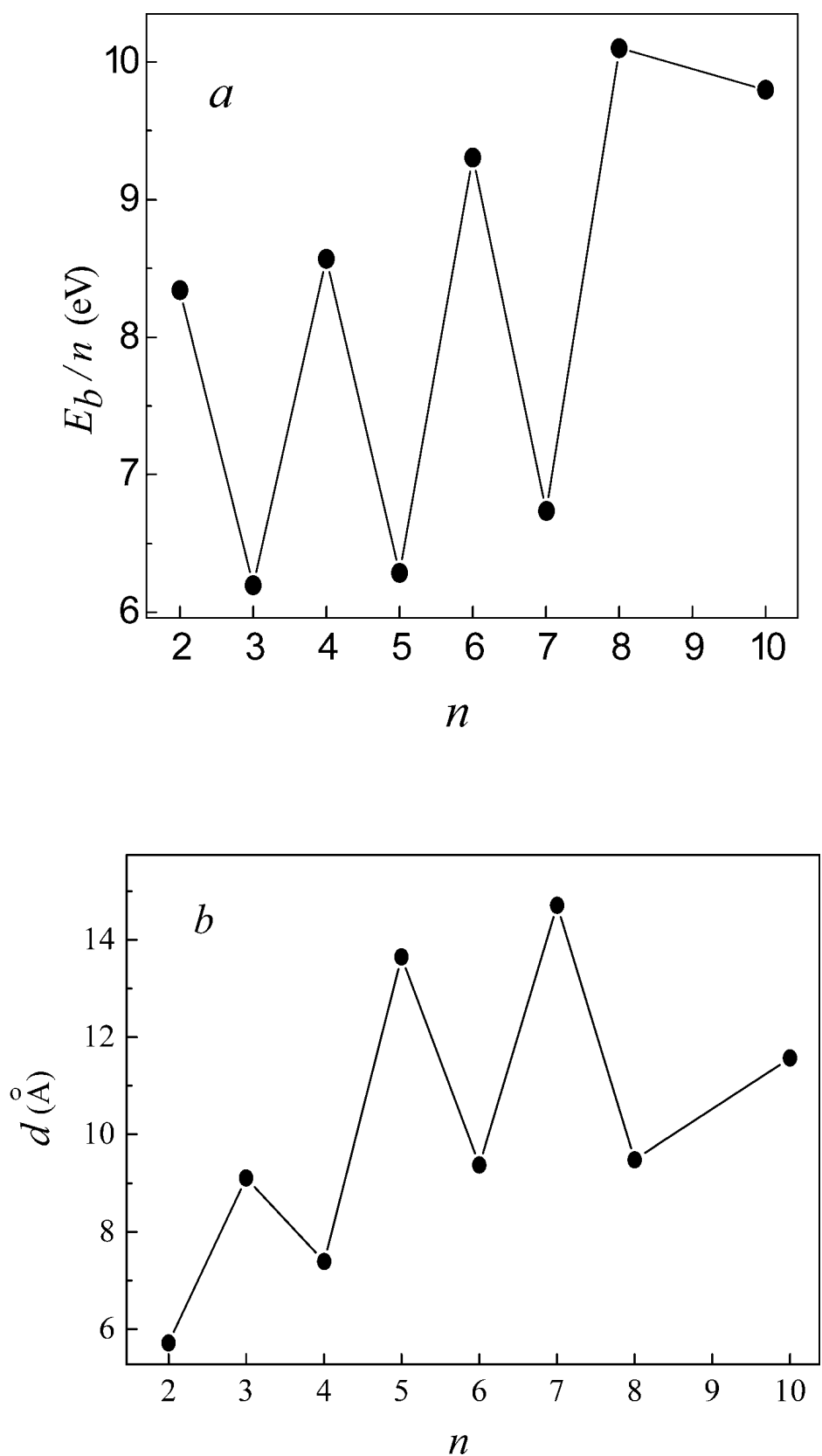

Fig.4. The results of the *ab initio* calculations of $(CdP_2)_n$ cluster binding energy per formula unit (*a*) and maximum diameter of cluster (*b*) versus *n*.

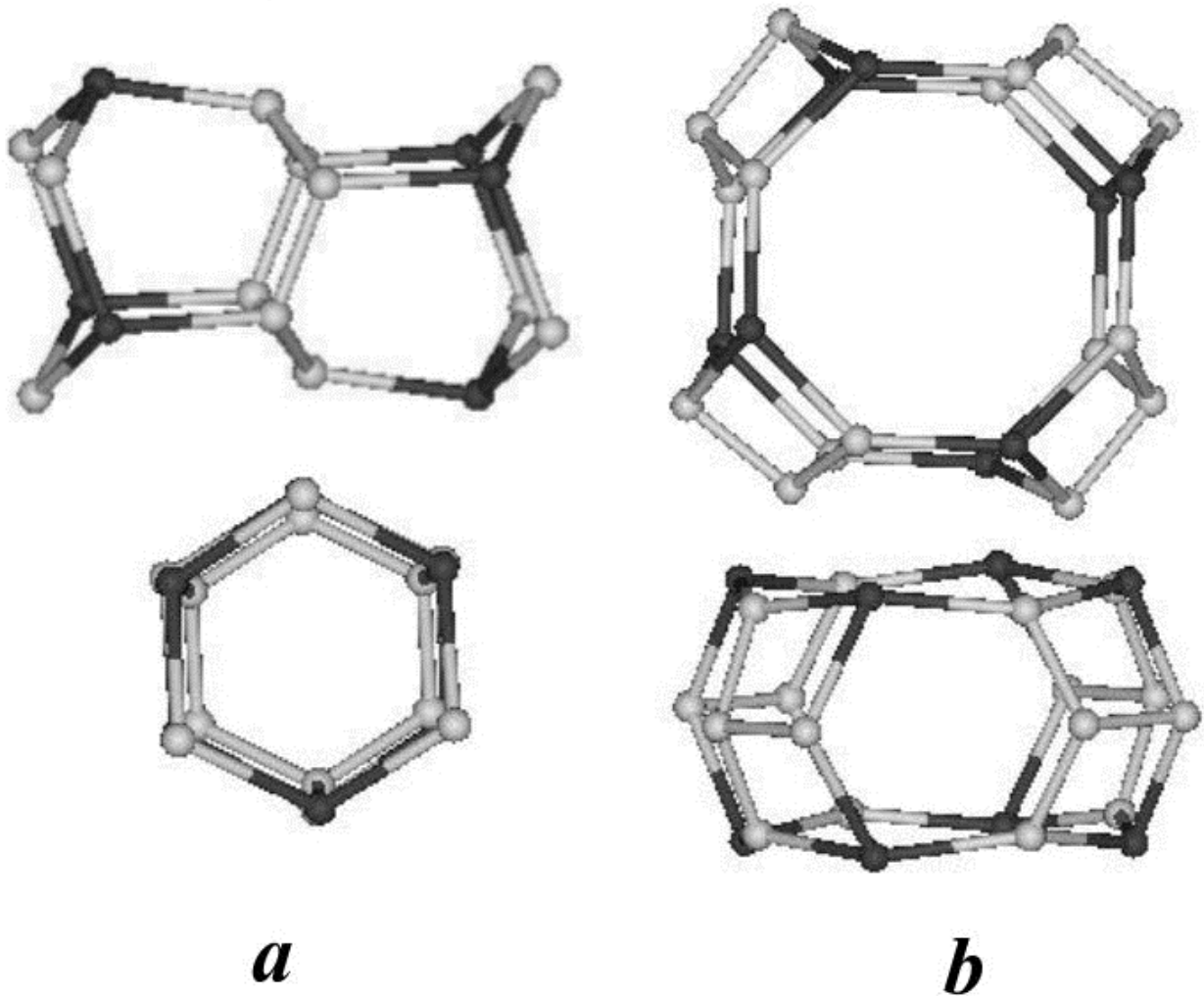

Fig.5. Calculated structure of the most stable $(CdP_2)_n$ clusters: (*a*) – structure of the $(CdP_2)_6$ cluster, and (*b*) – structure of the $(CdP_2)_8$ one, where Cd atoms – black balls, P atoms – grey balls.

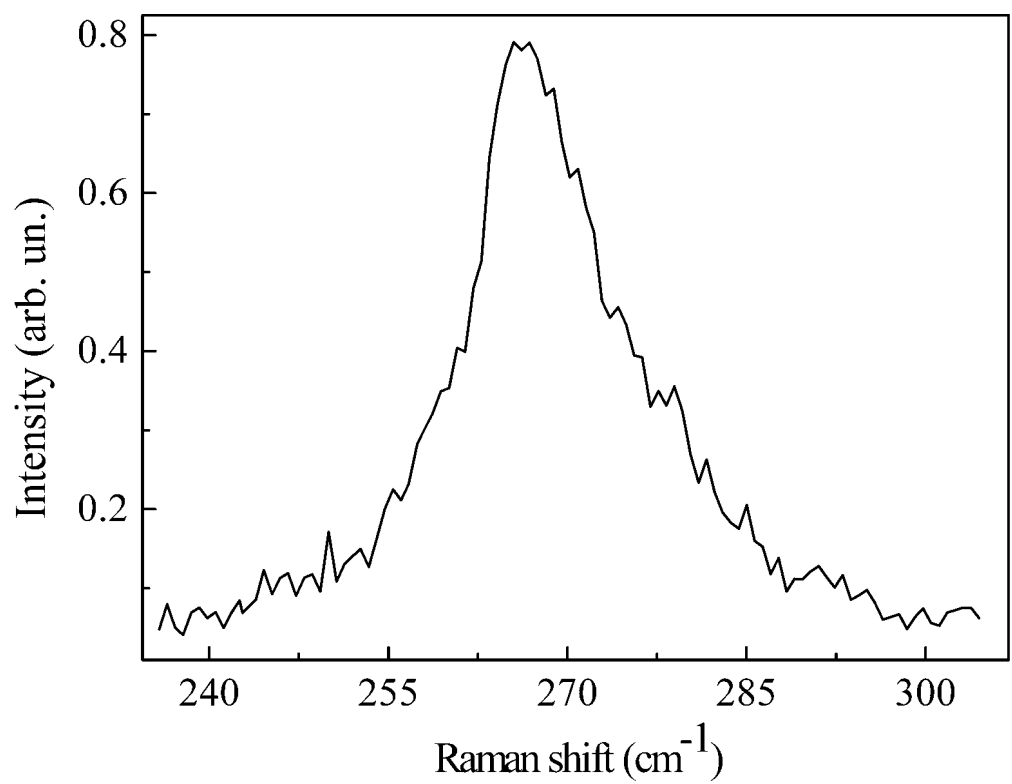

Fig.6. The Raman spectrum of $CdP_2$ clusters in zeolite Na-X at the temperature of 293 K.

Table 1. Spectral characteristics of $(CdP_2)_n$ clusters in zeolite Na-X and clusters produced by laser ablation.

| Spectral position (eV) | | | Blue shift of absorption band (eV) | Stokes shift (eV) | |
|---|---|---|---|---|---|
| Absorption | PL | | | Zeolite Na-X | Ablation |
| | Zeolite Na-X | Ablation | | | |
| 3.306 ($B_1$) | 2.532 ($B'_1$) | 2.267 ($B''_2$) | 1.151 ($B_1$) | 0.774 ($B_1$-$B'_1$) | 0.481 ($B_2$-$B''_2$) |
| 2.748 ($B_2$) | 2.264 ($B'_2$) | 2.071 ($B_b$) | 0.593 ($B_2$) | 0.484 ($B_2$-$B'_2$) | |